\documentstyle[aps,floats,psfig,prb,twocolumn]{revtex}
\draft

\begin{document}

\title{Bose
condensation of cavity polaritons beyond the linear regime: the
thermal equilibrium of a model microcavity.}

\author{P. R. Eastham and P. B. Littlewood} 
\address{Theory of Condensed Matter, Cavendish Laboratory, Cambridge, CB3 0HE.  United Kingdom.} 

\maketitle \begin{abstract} We consider a generalization of the Dicke
model. It describes localized, physically separated, saturable
excitations, such as excitons bound on impurities, coupled to a single
long-lived mode of an optical cavity. We consider the thermal
equilibrium of the model at a fixed total number of excitons and
photons. We find a phase in which both the cavity field and the
excitonic polarization are coherent. This phase corresponds to a Bose
condensate of cavity polaritons, generalized to allow for the
fermionic internal structure of the excitons. It is separated from the
normal state by an unusual reentrant phase boundary. We calculate the
excitation energies of the model, and hence the optical absorption
spectra of the cavity. In the condensed phase the absorption spectrum
is gapped. The presence of a gap distinguishes the polariton
condensate from the normal state and from a conventional laser, even
when the inhomogeneous linewidth of the excitons is so large that
there is no observable polariton splitting in the normal state.
\end{abstract} 

\section{Introduction}

In the strong-coupling regime for matter and light, radiative decay of
a material excitation gives way to coupled oscillations of the
polarization of the matter and of the electromagnetic field. The
quasiparticles corresponding to such coupled modes are known as
polaritons\ \cite{klingshirn-semiconopt}. The classic realization of
polaritons is excitons in a bulk semiconductor coupled to photons in
free space, as discussed many years ago by Hopfield\
\cite{hopfpol}. In this example, wavevector conservation ensures that
each exciton is coupled only to a single mode of the electromagnetic
field, leading to the formation of polaritons which are superpositions
of a single exciton and photon. Recently, there has been a lot of
interest in polaritons formed from photons confined in cavities: such
cavity polaritons have now been observed for confined photons coupled
to atoms\
\cite{atompol}, to two-dimensional excitons in quantum wells\
\cite{cavpol}, to bulk excitons\ \cite{bulkcavpol}, to excitons in
films of organic semiconductors\ \cite{organicpol1,organicpol2}, and
to charged exciton complexes\ \cite{chargedpol}.

Since polaritons are photons coupled to other excitations, they are
bosons, and so are candidates for Bose condensation\
\cite{keldyshbec}. Recent observations\
\cite{senbloch,cdteboser,bulkboser,angleboser,angle-cw-stimscat} of
bosonic behavior for cavity polaritons have renewed interest in this
idea.

However, there is a conceptual difficulty with a Bose condensate of
polaritons: while polaritons are usually considered in the
low-excitation linear regime, Bose condensates are stabilized by
nonlinearities\ \cite{nozieresbec}. For cavity polaritons, there is
also the following more practical difficulty. Bose condensates are
characterized by coherence, and in a polariton condensate this
coherence will appear in the photons. Given this, how is a polariton
condensate distinct, conceptually and observationally, from a laser?

In this paper we address these problems by developing a theory of
polariton condensation in the Dicke model\
\cite{dickemodel}. This nonlinear model of confined photons coupled to
matter is one of the basic models of laser physics. It allows us to go
beyond the conventional linear-response concept of a polariton,
including effects due to finite excitations of the matter in the
cavity. In the language of semiconductors, it includes a
``saturation'' or ``band-filling'' nonlinearity, produced by the
fermionic internal structure of the excitons.

Polaritons are not conserved particles, so there is ultimately no
equilibrium condensate. We may, however, treat polaritons as conserved
particles if their lifetime is much longer than the time required to
achieve thermal equilibrium at a fixed polariton number. We will study
this quasi-equilibrium regime, since it is in this regime that Bose
condensation is well-defined\ \cite{keldyshbec}.

In section\ \ref{modelsec}, we introduce the model, and explain how
the concept of a polariton can be generalized to allow for the
nonlinearity of the model. We then present, in sections\
\ref{variational} and\ \ref{variationalresults}, a simple variational
technique for calculating the ground state of the model at a fixed
density of polaritons. In section\
\ref{largensec} we investigate the thermodynamics
of the model using an alternative technique based on functional
integrals. This technique demonstrates that the variational approach
is essentially exact, and allows us to consider finite
temperatures. In section\
\ref{phasediag}, we use the expressions derived by the functional integral
method to study the phase diagram for condensation, while in section\
\ref{exspec}, we use these expressions to calculate the excitation
spectra of the model. These excitation spectra provide a physical
picture of the transition to the condensed state, and determine the
absorption spectrum of the cavity. Finally, in section\
\ref{conclusions} we discuss our conclusions.

The functional integral approach to the thermodynamics of our model
has already been the subject of a brief report\ \cite{letterpaper}. We
extend that earlier report to allow for a distribution of the energies
of the electronic excitations, i.e. inhomogeneous broadening, which is
significant in many potential realizations of the polariton
condensate.

\section{Model}
\label{modelsec}

The Dicke model\ \cite{dickemodel,alleneberly} consists of a set of
$N$ two-level oscillators coupled to a single mode of the
electromagnetic field by the dipole interaction. The two-level
oscillators do not interact with one another, except through their
common coupling to the electromagnetic field. We generalize the
original Dicke model to include an energy distribution of the
two-level oscillators. Making the rotating-wave approximation(see
e.g. Refs.\
\onlinecite{alleneberly,sculzub}), we consider the Hamiltonian
\begin{eqnarray}
\label{ham}
H = \sum \frac{E_{g}(n)}{2} \left( b^{\dagger}b-a^{\dagger}a
\right) + \omega_{c} \psi^{\dagger}\psi + H^{\prime}, \\
H^{\prime}= \frac{g}{\sqrt{N}}\sum \left( b^{\dagger} a \psi  +
\psi^{\dagger} a^{\dagger} b  \right). \nonumber
\end{eqnarray} Here the two-level oscillators are indexed by the
variable $n$, which is summed over. We use a fermionic representation
for the two-level oscillators, describing each one in terms of a pair
of fermions with annihilation operators $a$ and $b$. For brevity we
suppress the index $n$ on the fermionic operators. The fermions are
subject to the single-occupancy constraint
\begin{equation}
\label{constraints}
b^{\dagger} b + a^{\dagger} a =1 \end{equation} on each
site. $\psi$ is the annihilation operator for the cavity mode,
$E_{g}(n)$ is the energy of the $n^{th}$ two-level oscillator, and $g$
is the strength of the dipole coupling.

The Hamiltonian (\ref{ham}) is a simple model of a three-dimensional
cavity (photonic dot)\ \cite{pillarcavpol,cav3d}, containing
localized, physically separated electronic excitations. Although
simplified, it is a useful starting point for many systems. For
example, each of the two-level oscillators could describe the presence
or absence of a localized exciton in a given eigenstate of the
disorder potential in a disordered quantum well, on a given molecule
in an organic film, or trapped on a particular impurity. The
restriction to singly occupied states models the hard-core repulsion
produced by the fermionic structure of such excitations. It describes
spinless excitations which are localized on the scale of the (exciton)
Bohr radius. It is straightforward to generalize our calculations to
allow for a finite number of excitations on each site, i.e. traps
which are bigger than the Bohr radius, and so can hold several
excitons.

$H^\prime$ couples the photons to excitations of the two-level
oscillators, created by the operator $S_{+}=\frac{1}{\sqrt{N}}\sum
b^\dagger a$. If $E_{g}(n)=E_{g}$ then the excited states which are
created by $S_{+}$ from the vacuum are eigenstates of the bare
Hamiltonian $H$. If furthermore $N$ is large and the two-level
oscillators are near to their ground state,
\begin{equation} \frac{1}{2}\sum
\langle b^\dagger b -a^\dagger a \rangle \approx -N/2, \label{lowdensity-regime}
\end{equation} then $S_{+}$ is approximately a bosonic creation
operator, and the Hamiltonian (\ref{ham}) becomes two coupled boson
oscillators\footnote{This mapping may be formalized\ \cite{thesis}
using the Holstein-Primakoff transformation.}. Polaritons are usually
presented as the eigenstates of such a model.

Away from the low-excitation limit (\ref{lowdensity-regime}), $S_{+}$
is not a bosonic creation operator, and the conventional description
of polaritons breaks down. To go beyond the low-excitation limit, we
generalize the concept of a polariton to be the quantum of excitation
of the coupled matter-light system. The polariton number is then the
total number of photons and excited two-level oscillators,
\begin{eqnarray} N_{\mathrm{pol}}&=& L+N/2 \label{exnumber-definition}
\\ &=&\psi^{\dagger}
\psi+\frac{1}{2}\sum b^{\dagger}b-a^{\dagger}a + N/2, \nonumber \end{eqnarray}
which is a conserved quantity for the model
(\ref{ham}). (\ref{exnumber-definition}) defines the operator $L$,
which we refer to as the excitation number. We define a corresponding
excitation density $\rho_{ex}=\langle L \rangle/N$, which is the total
number of photons and electronic excitations, per two-level
oscillator, minus one-half. Since the numbers of photons and
electronic excitations are positive, the lowest excitation density is
$-0.5$. Since the number of electronic excitations is always less than
$N$, the electronic contribution to $\rho_{ex}$ is always less than
$0.5$.

The thermal equilibrium of the Dicke model, in the absence of an
externally created population of polaritons, has been studied
extensively since the pioneering exact solution of Hepp and Lieb\
\cite{hepplieb}. These authors showed that, even in the absence of
external excitation, the Dicke model has a phase transition to a Bose
condensed state. Such an equilibrium condensate is a static, coherent
state of photons: it is a ferroelectric\
\cite{dicke-ferroelectric}. Here we are interested in the thermal
equilibrium of a population of polaritons: the quasi-equilibrium
problem posed by (\ref{ham}) at a fixed excitation $L$. The
quasi-equilibrium condensate which we find in this regime is a
time-varying generalization of the ferroelectric state discovered by
Hepp and Lieb.

\section{Variational approach} 
\label{variational}

We can write down a variational state which describes the polariton
condensate by noting that Bose condensates are described by coherent
states. This produces a variational wavefunction closely related to
the BCS wavefunction used to describe superconductors and exciton
condensates. As has been stressed by Comte and Nozi\`{e}res\
\cite{nozieresex1}, this class of wavefunction can describe an exciton
condensate in both the low and high density limits. It thus permits a
smooth interpolation from low densities, where the excitons are simple
bosons, to high densities, where their fermionic internal structure is
revealed. In a similar way, it allows us to explore the polariton
condensate beyond the low-density regime in which polaritons are
usually considered.

The constituents of our proposed condensate are the excitations of the
whole system, counted by $L$. In general, such an excitation is a
superposition of an excitation of the cavity mode and an excitation of
the electronic states. Thus we take for our trial wave-function a
coherent state of such a superposition:
\begin{equation} |\lambda,w\rangle=e^{\lambda \psi^\dagger + \frac{1}{\sqrt{N}}\sum_{n} w_{n}
b^{\dagger}a}|vac\rangle,
\label{becform}
\end{equation} where the state $|vac\rangle$ has a single fermion in
the lower state of each two-level oscillator. The state
(\ref{becform}) has a finite polarization of the electronic
excitations as well as a finite amplitude for the cavity
field. $\lambda$ and $w_{n}$ are the variational parameters. Expanding
the exponential, (\ref{becform}) explicitly becomes a superposition of
a coherent state of photons and a BCS state of the fermions,
\begin{equation}
\label{varstate} |\lambda,u,v\rangle = e^{\lambda \psi^\dagger }
\prod_{n} ( v_{n} b^{\dagger} + u_{n}e^{i\phi_{n}} a^{\dagger} )
|0\rangle.
\end{equation} Here $\lambda,
u_{n}, v_{n}$, and $\phi_{n}$ are the variational parameters, and
$|0\rangle$ denotes the vacuum state with no fermions in any of the
levels. By construction, this variational state obeys the
single-occupancy constraints (\ref{constraints}). We fix the overall
phase of the condensate by choosing $\lambda$ to be real. The
$\phi_{n}$ have been explicitly introduced to make the $u$ and $v$
real. They are the phase differences between the cavity field and the
polarizations of the electronic states.

To find the ground state of (\ref{ham}) at fixed excitation number we
minimize
\begin{eqnarray}
\label{hamexps}
 \langle H-\mu_{ex}L\rangle &=&\tilde{\omega}_{c} \lambda^2 +
\sum_{n} \tilde{\varepsilon}_{n} (v_{n}^2-u_{n}^2) \nonumber \\
& & + 2\frac{g}{\sqrt{N}}\lambda u_{n}v_{n} \cos(\phi_{n}), \\
\tilde{\omega}_{c} & = & \omega_{c}-\mu_{ex}, \nonumber \\
\tilde{\varepsilon}_{n} & = & \frac{E_{g}(n) - \mu_{ex}}{2}, \nonumber
\end{eqnarray} with respect to the variational parameters, subject to 
the normalization conditions $u^2_{n}+v^2_{n}=1$.

Although the overall phase of the condensate is arbitrary, the
relative phases $\phi_{n}$ are not: there is only one order
parameter. The relative phases $\phi_{n}$ are fixed by the last term
in (\ref{hamexps}), the dipole coupling. This term ensures that all
the two-level oscillators which have a finite dipole moment($u_{n}\neq
0, 1$) are mutually coherent, $\phi_{n}=\phi$, when the energy is
minimized. It is the dipole interaction which is responsible for Bose
condensation, and its accompanying coherence\ \cite{nozieresbec}, in
the present system.

Setting $\phi_{n}=0$ and defining an intensive $\lambda$ by rescaling
$\lambda\to \lambda\sqrt{N}$, the condensate parameters are given by
the real solutions with $\lambda u_{n} v_{n} < 0$ to
\begin{eqnarray} \label{extreqs}
\tilde{\omega}_{c} \lambda + \frac{g}{N}\sum_{n}u_{n} v_{n} =0, \\
2\tilde{\varepsilon}_{n} u_{n} v_{n} - g \lambda (v_{n}^2-u_{n}^2) =
0. \nonumber 
\end{eqnarray}

$\mu_{ex}$ was introduced as a Lagrange multiplier constraining the
excitation number. It is the chemical potential for our coupled modes,
and is related implicitly to the excitation density by
\begin{eqnarray} \label{density}
\rho_{ex} &=& \frac{1}{N} \left\langle \psi^{\dagger}\psi +
\frac{1}{2} \sum b^{\dagger}b-a^{\dagger}a
\right\rangle \nonumber \\ &=& \lambda^2 + \frac{1}{2N} \sum_{n}
v^{2}_{n}-u^{2}_{n}.
\end{eqnarray}

Eliminating $u_{n}$ and $v_{n}$ from (\ref{extreqs}) and
(\ref{density}) we can rewrite these expressions as
\begin{eqnarray}
\tilde{\omega}_{c} \lambda=\frac{g^2
\lambda}{2N}\sum_{n}\frac{1}{|E_{n}|},
\label{ztextr} \\
\rho_{ex}=\lambda^2-\frac{1}{2N}\sum_{n}\frac{\tilde\varepsilon_{n}}{|E_{n}|},
\label{ztexdens}
\end{eqnarray} where we define \begin{equation}
\label{qpenergy}
E_n=\mathrm{sign}(\tilde{\varepsilon}_{n})
\sqrt{\tilde{\varepsilon}^2_n+g^2|\lambda|^2}.\end{equation} (\ref{ztextr}) is
analogous to the BCS gap equation, with an order parameter $\lambda$.

\section{Zero temperature properties}
\label{variationalresults}

To investigate the expressions (\ref{ztextr}--\ref{qpenergy}), we
replace the summations over sites with an integral over the energy
distribution of the two-level oscillators. We take this distribution
to be a Gaussian with mean $E_{0}$ and variance $\sigma g$. The
remaining parameters in our quasi-equilibrium problem are then the
excitation density, $\rho_{ex}$, and the dimensionless detuning
between the energy of the cavity mode and the center of the exciton
line, $\Delta=(\omega_{c}-E_{0})/g$.

For a Gaussian density of states, the summation on the right of
(\ref{ztextr}) diverges as $\lambda\to 0$, and approaches zero as
$\lambda \to \infty$. Thus for any $\mu_{ex}<\omega_c$ there is always
a condensed solution, $\lambda\neq0$, to (\ref{ztextr}): the system is
condensed at arbitrarily small excitation densities. This behavior is
produced by the tails of the Gaussian distribution. Because of these
tails, we have excitons at arbitrarily low energies, and hence also
bound exciton-photon states at arbitrarily low energies. It is
impossible to populate just the excitons, because no matter how small
$\mu_{ex}$ is, there is always a bound state involving photons below
it. We expect that if the density of states has a lower cut-off, and
is continuous at this cut-off, there would be a finite critical
$\mu_{ex}$ below which there is no condensed solution to
(\ref{ztextr}).

\begin{figure}[t]
\centerline{\psfig{file=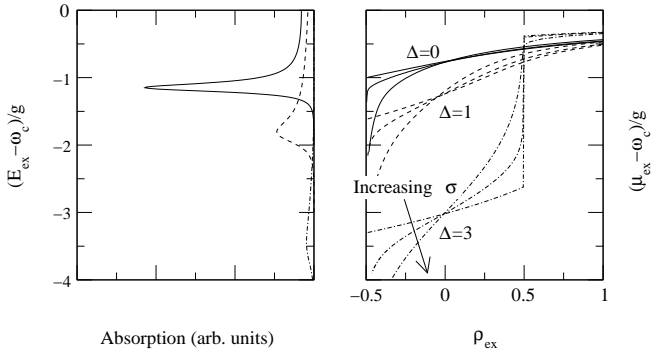,width=8.6cm}}
\caption{Right panel: dependence of the chemical potential on
excitation density for detunings $\Delta=0, 1$ and $3$ and variances
$\sigma=0, 0.5$ and $1$. Left panel: absorption spectrum for a
microcavity at $\rho_{ex}=-0.5$ and $T=0$ for $\sigma=0.5$ and the
same three detunings.}
\label{zerotmurhofig}
\end{figure}

Let us investigate the dependence of $\mu_{ex}$ on $\rho_{ex}$ in the
absence of inhomogeneous broadening, $\sigma=0$. At low densities,
$\rho_{ex}\approx -0.5$, $\mu_{ex}$ can be obtained from
(\ref{ztextr}). Expanding this expression for small $\lambda$ and
comparing the leading terms, we find that $\mu_{ex}$ is given by the
conventional linear-response polariton energy, $\mu_{ex}=E_{\mathrm{
LPB}}=\frac{1}{2}[(\omega_{c}+E_{g})-g\sqrt{\Delta^2+4}]$. At finite
densities we calculate $\mu_{ex}$ numerically, by solving
(\ref{ztextr}) and (\ref{ztexdens}) to determine
$\rho_{ex}(\mu_{ex})$. The results are plotted in the right panel of
Fig.\ \ref{zerotmurhofig}, for $\Delta=0, 1$ and $3$. At low densities
we are describing a condensate of conventional polaritons, and so have
$\mu_{ex}=E_{\mathrm{LPB}}$. As the density is increased the exciton
states saturate, forcing the excitations to become more
photon-like. Thus the chemical potential approaches $\omega_{c}$ at
high densities. For $\Delta > 2$ the separation between the
exciton-like and photon-like excitations persists to $\rho_{ex}=0.5$,
where the exciton states are completely saturated. This results in a
discontinuity in $\mu_{ex}$ at this point, since no further excitation
can be added to the exciton states.

The dependence of $\mu_{ex}$ on $\rho_{ex}$ in the inhomogeneously
broadened case is also illustrated in the right panel of Fig.\
\ref{zerotmurhofig}. It is qualitatively rather similar to the homogeneous
case. Instead of the finite intercept of the homogeneous case we now
have $\mu_{ex}\to-\infty$ as $\rho_{ex}\to -0.5$. This behavior is
again caused by the tails of the Gaussian distribution. To demonstrate
how $\mu_{ex}$ approaches the conventional polariton energy
$E_{\mathrm{LPB}}$ in the homogeneous, low-density limit, we compare
the behavior of $\mu_{ex}$ with the density of states for the
linear-response excitations of the empty($\rho_{ex}=-0.5$)
cavity. This density of states is the optical absorption spectrum of
the cavity, and is plotted in the left panel of Fig.\
\ref{zerotmurhofig} for $\sigma=0.5$ and $\Delta=0, 1$ and $3$. We
will describe how it is calculated in section\ \ref{exspec}. At very
low densities, $\mu_{ex}$ lies in the tails of the exciton
distribution. With increasing density, these states quickly saturate,
producing a sharp rise in $\mu_{ex}$. As $\mu_{ex}$ reaches the
polariton peak, the sharp rise in the density of states for the
coupled modes produces a kink in the chemical potential. In the
homogeneous limit, this kink moves to zero density and corresponds to
the usual polariton energy. Since the density of states at this point
is infinite in the homogeneous limit, these polaritons are simple
bosons.

\begin{figure}[t]
\centerline{\psfig{file=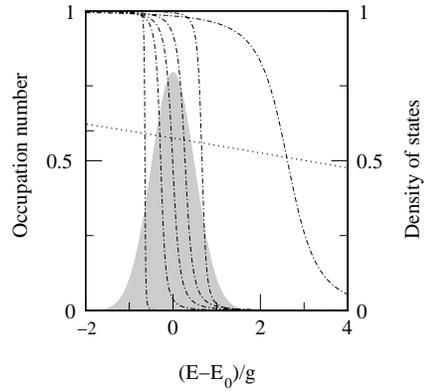,width=5.5cm}}
\caption{Occupation of the two-level oscillators at zero temperature
as a function of energy E for $\Delta=3$, $T=0$, $\sigma=0.5$ and
densities $\rho_{ex}=-0.4, -0.2, 0, 0.2, 0.4, 0.6$(dot-dashed curves,
increasing from left to right) and $\rho_{ex}=100$(dotted curve). The
shaded region shows the Gaussian distribution of oscillator energies
used.}
\label{zerotoccfig}
\end{figure}

Figure\ \ref{zerotoccfig} shows the occupation of the two-level
oscillators in the polariton condensate, for $\Delta=3, \sigma=0.5$
and various densities. The occupation number of the $n^{th}$ two-level
oscillator is \begin{displaymath} \frac{1}{2}(v_{n}^2-u_{n}^2+1)=
\frac{1}{2}\left(1-\frac{\tilde{\varepsilon}_{n}}{|E_{n}|}\right).\end{displaymath}
As is clear from the figure, this is a Fermi step broadened by the
interaction with the photons, just as the electronic distribution in a
BCS superconductor is a Fermi step broadened by the pairing
interaction. The states in the broadened region of the step have a
finite dipole moment and are involved in the condensate. The Fermi
step moves up through the exciton line as the excitation is increased
from $\rho_{ex}=-0.5$ and the low-lying electronic states saturate. At
very large densities there are a large number of photons, and the
Fermi step is almost completely flat: rather than the electronic
system completely saturating in the high density limit, it approaches
half filling. This is because the half-filled state maximizes the
polarization of the electronic states and hence minimizes the dipole
interaction between the excitons and the macroscopically occupied
cavity mode.

\begin{figure}[t]
\centerline{\psfig{file=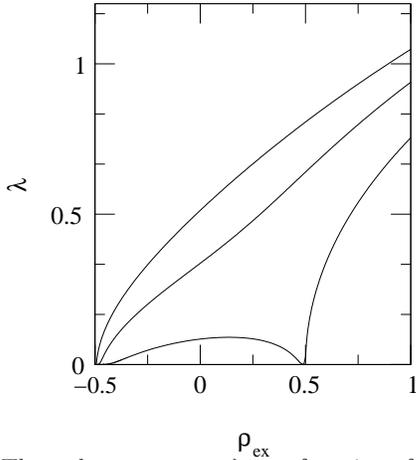,width=5.5cm}}
\caption{The order parameter $\lambda$ as a function of density, for
$\sigma=0.5$ and $\Delta=0$, 1 and 3. $\lambda^2$ is the photon number
per two-level oscillator in the condensed state.}
\label{photonoccfig}
\end{figure}

Careful inspection of Fig.\ \ref{zerotoccfig} reveals that the
broadening of the Fermi step produced by the photons does not increase
monotonically with density. This corresponds to a non-monotonic
dependence of the field amplitude, $\lambda$, on density. This
dependence is illustrated in Fig.\
\ref{photonoccfig}. The field amplitude is related to the
electronic polarization by the first of the equations
(\ref{extreqs}). It is proportional to the electronic polarization and
inversely proportional to the separation between the chemical
potential and the cavity mode. The electronic polarization depends on
the density of states in the vicinity of the chemical potential(Fig.\
\ref{zerotoccfig}); the peak in the density of states at the center of the
exciton line produces the peak in Fig.\
\ref{photonoccfig}.

\section{Large-N Expansion} 
\label{largensec}

The variational approach of sections\ \ref{variational} and \
\ref{variationalresults} becomes exact in the thermodynamic limit
$N\to\infty$. Physically, this is because it corresponds to a
mean-field treatment of the interaction between electronic
excitations. This interaction, between a large number ($N$) of
electronic excitations, is mediated by a small number (one) of cavity
modes. In a mean-field treatment of this interaction, each electronic
excitation is coupled to the average field produced in the cavity by
the other electronic excitations. This becomes exact when there are a
large number of electronic excitations contributing to a small number
of field modes, since the fluctuations of the field are then
negligible.

In this section, we develop a mean-field theory for the thermodynamics
of the model (\ref{ham}) from the functional-integral representation
of the partition function. In this representation, the partition
function can be rigorously evaluated, for large $N$, using a
saddle-point analysis\ \cite{pybook}. From such an analysis, we derive
finite-temperature generalizations of the variational expressions
(\ref{ztextr}--\ref{qpenergy}), thus demonstrating that they are
rigorous in the limit of large $N$.

The functional integral techniques used here have previously been
used\ \cite{pybook,leemodel-pi} to calculate the partition function
and excitation energies {\emph{in the absence of a constraint on the
polariton number}} of a simplification of the Dicke model. While the
Hamiltonian of the model discussed in Refs.\
\onlinecite{pybook,leemodel-pi} is given by (\ref{ham}) with
$E_{g}(n)=E_{g}$, the local constraints prohibiting two fermions on
the same site, (\ref{constraints}), are replaced with a global
constraint. In contrast, we retain (\ref{constraints}) as local
constraints, as well as including a distribution of $E_{g}$ and a
constraint on the polariton number.

As in sections\ \ref{variational} and \ \ref{variationalresults}, we
work in a grand-canonical ensemble, using a chemical potential
$\mu_{ex}$ to constrain the excitation number. We consider the
partition function associated with this ensemble \begin{displaymath} Q
= {\mathrm{Tr}}\, e^{-{\beta(H-\mu_{ex}L)}}.
\end{displaymath} The coherent-state functional-integral formalism
allows us to express $Q$, for the model (\ref{ham}), as the
constrained functional integral
\begin{displaymath}
Q=\int {\cal D}\psi \prod_n [ {\cal
D}\eta_n\delta(\bar{\eta}_n\eta_n-1)]e^{-S},
\end{displaymath}
with the action
\begin{displaymath}
S=\int_{0}^{\beta}d\tau
\bar{\psi}(\partial_{\tau}+\tilde{\omega}_{c})\psi + \sum_n
\bar{\eta}_n M_{n}\eta_{n}. \end{displaymath}
We have introduced a Nambu spinor \begin{displaymath}
\eta_{n}=\left( \begin{array}{c}
b_{n} \\ a_{n}
\end{array}
\right) \end{displaymath} for each two-level oscillator. The matrix $M_n$ is \begin{displaymath} M_{n}=\left(
\begin{array}{cc}
\partial_{\tau}+\tilde{\varepsilon}_{n} & g\psi/\sqrt{N} \\
g\bar{\psi}/\sqrt{N} & \partial_{\tau}-\tilde{\varepsilon}_{n}
\end{array} \right).
\end{displaymath} 

Rescaling the boson field $\psi\to\sqrt{N}\psi$ and transferring the
fermionic integrals into the action gives
\begin{displaymath}
Q=\int{\cal D}\psi |J| e^{-N S_{\rm eff}},
\end{displaymath}
with an effective action
\begin{eqnarray}
S_{\rm eff}&=& \int_{0}^{\beta}d\tau 
\bar{\psi}(\partial_{\tau}+\tilde{\omega}_{c})\psi - \frac{1}{N}\sum_{n}S_{{\rm f},n},
\label{seff}
\\ S_{{\rm f},n}&=&\ln \int {\cal
D}\eta_{n}\delta(\bar{\eta}_{n}\eta_{n}-1)e^{-\int_{0}^{\beta}\bar{\eta}_{n}P_{n}\eta_{n}},
\nonumber
\end{eqnarray} in which the $P_{n}$ are the matrix operators $M_{n}$ after rescaling the
boson field, and $J$ denotes the trivial Jacobian arising from this
rescaling.

\subsection{Mean-field equation}

For large $N$, the dominant contribution to the partition function $Q$
comes from those functions $\psi_{0}(\tau)$ which minimize the action
$S_{\mathrm{eff}}$. Such functions obey the Euler-Lagrange
equation. For the action (\ref{seff}), this takes the
form\begin{eqnarray}
(\partial_{\tau}+\tilde{\omega}_{c})\psi_{0}(\tau)&=&\frac{1}{N}\sum_{n}\left.\frac{\delta
S_{{\rm f},n}}{\delta\bar{\psi}}\right|_{\psi(\tau)=\psi_{0}(\tau)}
\nonumber \\
&=&-\frac{g}{N}\sum_{n}\langle\bar{a}_{n}(\tau)b_{n}(\tau)\rangle,
\label{geneleq}\end{eqnarray} where the right-hand side of this
expression is the polarization of the two-level oscillators in thermal
equilibrium driven by an external field $\psi_{0}(\tau)$. This
polarization appears because the field $\psi_{0}(\tau)$ modifies the
eigenstates\ \cite{gelesin} of the electronic system. A thermal
population of these new eigenstates can correspond to a finite
polarization of the original fermions. Equation (\ref{geneleq}) is a
self-consistency condition: the cavity field is driven by the
polarization of the fermions, which itself arises from the
renormalization of the fermions produced by the photons.

Assuming that the self-consistent field $\psi_{0}(\tau)$ is
independent of $\tau$, we can calculate the polarization term on the
right of (\ref{geneleq}) by making a Bogolubov transformation
\begin{equation} \label{bogt} \eta_{n} = \left(
\begin{array}{cc} \cos(\theta)e^{i\phi} & -\sin(\theta) \\
\sin(\theta) & \cos(\theta)e^{-i\phi} \end{array} \right) \left(
\begin{array}{c} \delta_{n} \\ \gamma_{n} \end{array}
\right) \end{equation} from the $b_n$ and $a_n$ fermions to new
fermions $\delta_{n}$ and $\gamma_{n}$. This transformation
diagonalizes $P_{n}$ when $\phi= \arg \lambda$ and $\tan
2\theta=g|\lambda|/\tilde{\varepsilon}_{n}$. The $\delta_n$ and
$\gamma_n$ quasiparticles then have energies $\pm E_{n}$ respectively,
with $E_{n}$ defined by equation (\ref{qpenergy}). Since (\ref{bogt})
is a rotation in $\eta$ space, it preserves the single occupancy
constraints. Thermally populating the new fermions in accordance with
the single occupancy constraint we have
\begin{eqnarray}
\langle
\bar{a}_{n}b_{n}\rangle&=&\frac{1}{2}e^{i\phi}\sin(2\theta)\langle\bar{\delta}_n
\delta_{n}-\bar{\gamma}_{n}\gamma_{n}\rangle
\nonumber \\ &=&
\frac{1}{2}e^{i\phi}\sin(2\theta)\tanh(\beta E_n), \nonumber
\end{eqnarray}
and (\ref{geneleq}) becomes
\begin{equation}
\tilde{\omega}_{c} \lambda=\frac{g^2
\lambda}{2N}\sum_{n}\frac{1}{E_{n}}\tanh\left(\beta E_{n}\right).
\label{ftextr}
\end{equation} 

Equation (\ref{ftextr}) is the finite-temperature generalization of
the variational result (\ref{ztextr}). This generalization is rather
straightforward: we have just acquired $\tanh(\beta E)$ factors
describing the thermal occupation of the two-level oscillators.

If we remove the constraint on the polariton number, by setting
$\mu_{ex}=0$, and set $E_{g}(n)=E_{g}$, then (\ref{ftextr}) is the
form originally derived by Hepp and Lieb\ \cite{hepplieb} for the
unconstrained equilibrium of the Dicke model. In that problem, the
existence of a condensate requires
\begin{equation} \frac{\omega_{c}E_{g}}{g^2} < 1, \label{minimumg}
\end{equation} since otherwise (\ref{ftextr}), with $\mu_{ex}=0$ and
$E_{g}(n)=E_{g}$, has only the trivial solution $\lambda=0$. However,
it is shown in Refs.\
\onlinecite{dickea2term1} and \onlinecite{dickea2term2} that the $A^2$ terms of the minimal-coupling Hamiltonian, neglected in the model (\ref{ham}), modify the
inequality\ (\ref{minimumg}) in a way which is inconsistent with the
Thomas-Kuhn-Reich sum rule. This sum rule requires $\kappa E_{g}/g^2 >
1$, where $\kappa$ is the coupling constant for the $A^2$ term, while
the modified inequality (\ref{minimumg}) reads
\begin{equation}
\frac{(\omega_{c}+2\kappa)E_{g}}{g^2} < 1. \label{minimumg2}
\end{equation} Since this inequality cannot be satisfied, the phase transition in the unconstrained case is an unphysical
artifact of the model (\ref{ham}). However, we do not believe that the
$A^2$ terms prevent condensation in the constrained case, because the
inequality corresponding to (\ref{minimumg2}) will be
\begin{displaymath}
\frac{(\tilde{\omega}_{c}+2\kappa)(E_{g}-\mu_{ex})}{g^2} <
1, 
\end{displaymath} and the parameter $\mu_{ex}$ is not restricted by the sum rule.

\subsection{Effect of fluctuations}
\label{fluc-nocontrib-subsec}

Let us now consider the effect of small fluctuations
$\delta\psi(\tau)$ around the mean-field solution. Expanding $S_{\rm
eff}$ to second order in a functional Taylor series around the
mean-field solution we have
\begin{equation} Q \approx e^{-NS_{0}}\int {\cal D}(\delta\psi)|J|
e^{-NS_{2}[\delta\psi,\bar{\delta\psi}]}. \label{gaussian-flucs}
\end{equation} Here $S_{0}$ is the action evaluated on the extremal
trajectory and $S_{2}$ is the quadratic action from the second order
term in the Taylor series. $S_{2}$ is the effective action for small
fluctuations of the electromagnetic field. The kernel of $S_{2}$,
${\mathcal{G}}^{-1}$, is the inverse of the thermal Green's function
for the photons.

The integral over fluctuations in (\ref{gaussian-flucs}) contributes a
term \begin{displaymath}\frac{1}{N}\ln
\det {\mathcal{G}}^{-1}\end{displaymath} to the free energy density. Since the
mean-field solution should be a minimum of the action, the eigenvalues
of ${\mathcal{G}}^{-1}$ should be positive. Then $\ln
\det {\mathcal{G}}^{-1}$ is finite as $N\to\infty$, there is no fluctuation
contribution to the free energy density in this limit, and the
mean-field theory becomes exact.

\subsection{Effective action for fluctuations} 
\label{effecac-subsec}

However, we have yet to check whether the solutions to (\ref{ftextr})
are actually minima of the action or merely extrema, i.e. whether the
mean-field solutions are stable against fluctuations. To check this,
we will need the effective action $S_{2}$, which we derive in this
section.

To obtain $S_{2}$, we calculate the (functional) second derivatives of
$S_{\rm eff}$, and evaluate them on the extrema
$\psi(\tau)=\psi_{0}(\tau)=\lambda$. In the frequency representation,
the components of $S_{2}$ are
\begin{eqnarray}\label{qpseffinhomgeneral1} \frac{\partial^{2}S_{\rm
eff}}{\partial{\psi(\omega)}\partial{\bar{\psi}(\omega^{\prime})}}=&&\beta
\delta(\omega^{\prime}-\omega)\bigg[i\omega + \tilde{\omega}_{c} \\ &&
-\frac{g^2}{N}\sum_{n}\int_{0}^{\beta} e^{-i\omega\tau} (\langle
\sigma_{n}^{-}(\tau)\sigma_{n}^{+}(0) \rangle \nonumber \\ &&
\phantom{-\frac{g^2}{N}\sum_{n}\int_{0}^{\beta} e^{-i\omega\tau} (}
-\langle \sigma_{n}^{-} \rangle\langle \sigma_{n}^{+}\rangle)\bigg]
d\tau, \nonumber
\end{eqnarray} and  \begin{eqnarray}\label{qpseffinhomgeneral2}
\frac{\partial^{2}S_{\rm{eff}}}{\partial{\psi(\omega)}\partial{\psi(\omega^{\prime})}}=&&-\beta
g^{2}\delta(\omega^{\prime}+\omega) \\ && \times
\frac{1}{N}\sum_{n}\int_{0}^{\beta}e^{i \omega
\tau}(\langle\sigma_{n}^{+}(\tau)\sigma_{n}^{+}(0)\rangle \nonumber \\
&& \phantom{\times \sum_{n}\int_{0}^{\beta}e^{i \omega
\tau}(}-\langle\sigma_{n}^{+}\rangle\langle\sigma_{n}^{+}\rangle)
d\tau. \nonumber
\end{eqnarray} $\omega$ and $\omega^\prime$ denote
bosonic Matsubara frequencies, $\sigma_{n}^{+}=b^{\dagger}_{n}a_{n}$
is the polarization operator for the $n^{th}$ two-level oscillator,
and the integrands are the susceptibilities of the two-level
oscillators in the self-consistent field $\lambda$. 

(\ref{qpseffinhomgeneral1}) and (\ref{qpseffinhomgeneral2}) describe
coupled fluctuations of the cavity field and the electronic
polarization. They are analogous to the Dyson-Gor'kov-Beliaev
equations\ \cite{pybook} of the theory of superconductors and
weakly-interacting Bose gases. Because we are working with a condensed
system there is an anomalous contribution to $S_{2}$,
(\ref{qpseffinhomgeneral2}), from fluctuations which do not conserve
the number of excitations above the condensate, i.e. those in which an
excitation enters or leaves the condensate.

To calculate the susceptibilities which appear in
(\ref{qpseffinhomgeneral1}), we rewrite them in terms of the
renormalized two-level oscillators using the transformation
(\ref{bogt}), transform to the Schr\"{o}dinger representation, and
take thermal and quantum-mechanical averages over the renormalized
eigenstates. This gives \begin{eqnarray}
\label{qpseffinhom} S_{2}[\delta\psi,\bar{\delta\psi}]&=&\beta \sum_{\omega}
\left(
\begin{array}{cc} \bar{\delta\psi}(\omega) & \delta\psi(-\omega)
\end{array} \right){\mathcal{G}}^{-1}\left( \begin{array}{c} \delta\psi(\omega) \\
\bar{\delta\psi}(-\omega) \end{array} \right), \nonumber \\
{\mathcal{G}}^{-1}&=&\left(\begin{array}{cc} K_{1} & K_{2} \\ K_{2}^{\ast} &
K_{1}^{\ast} \end{array} \right), \label{thermalinvgf} \\
K_{1}&=&i\omega+\tilde{\omega}_{c}+\frac{g^2}{N}\sum_{n}\bigg[\frac{1}{E_{n}}\tanh
\left(\beta E_{n}\right) \nonumber \\ && \times
\frac{i\tilde{\varepsilon}_{n}\omega-2\tilde{\varepsilon}_{n}^2-g^2|\lambda|^2}{\omega^2+4E_{n}^2}+\delta_{\omega}\alpha_{n}
|\lambda|^2 g^2 \bigg], \nonumber \\
K_{2}&=&\frac{g^4\lambda^2}{N}\sum_{n}\bigg[\frac{1}{E_{n}(\omega^2+4E_{n}^2)}\tanh\left(\beta
E_{n}\right) \nonumber \\ && +\delta_{\omega} \alpha_{n} \bigg], \nonumber \\
\alpha_{n}&=&-\frac{\beta}{4 E_{n}^2} {\rm{sech}}^2\left(\beta
E_{n}\right). \nonumber
\end{eqnarray} Note that, for the condensed state,
${\mathcal{G}}^{-1}$ takes different forms at $\omega=0$ and at finite
$\omega$. This is because thermal fluctuations at $\omega=0$ include
both fluctuations of the order parameter and quasiparticle
excitations\ \cite{hydroflucbook}. Only the latter appear at finite
$\omega$. In the normal state, $\lambda=0$, the effective action
simplifies to
\begin{eqnarray}\label{qpseff0}
S_{2}&=&\beta \sum_{\omega} \bar{\delta\psi}
(\omega)\Bigg[i\omega+\tilde{\omega}_{c} \nonumber \\ && + \frac{1}{N}\sum_{n}\frac{g^2 i}{\omega-2i\tilde{\varepsilon}_{n}}\tanh
\left(\beta\tilde{\varepsilon}_{n}\right)\Bigg]\delta\psi (\omega).
\end{eqnarray}

\subsection{Nature of the extrema}

We now use the expressions (\ref{qpseffinhom}--\ref{qpseff0}) to
investigate the nature of the extrema when $E_{g}(n)=E_{g}$.

Considering first a condensed solution, $\lambda\neq0$, we use the
extremal equation (\ref{ftextr}) to eliminate
$\frac{1}{E_{n}}\tanh\left(\beta E_{n}\right)$ from the matrix
${\mathcal{G}}^{-1}$. The eigenvalues of the resulting matrix are all
strictly positive provided that $\tilde{\omega}_{c}>0$, except for a
single zero eigenvalue at $\omega_{n}=0$. From (\ref{ftextr}) we see
that the condensed solutions always have $\tilde{\omega}_{c}>0$. Thus
we conclude that, at a condensed solution, the action has a minimum in
all but one direction, and is locally flat in this one direction.

We show in Appendix\ \ref{goldstappendix} that the single zero
eigenvalue describes a change in the overall phase of the
condensate. It is the Goldstone mode corresponding to the broken gauge
symmetry of the condensate. Because we are considering a broken
symmetry state, we should not integrate over these fluctuations when
calculating the partition function. Since the other eigenvalues of
${\mathcal{G}}^{-1}$ are always positive for the condensed solutions,
these solutions are stable against physical fluctuations, and the
mean-field theory is exact\footnote{In Appendix\ \ref{goldstappendix},
we give a formal demonstration that the zero mode does not contribute
to the free energy density as $N\to\infty$, so that the presence of
the zero mode does not invalidate the discussion of subsection\
\ref{fluc-nocontrib-subsec}.}.

Turning now to the normal solution, $\lambda=0$, we find from
(\ref{qpseff0}) that this is a minimum of the action unless
\begin{equation}\label{phaseboundinstab}
\tilde{\omega}_{c}<\frac{g^2}{2\tilde{\varepsilon}}\tanh(\beta\tilde{\varepsilon}).\end{equation}
This is just the condition for the extremal equation (\ref{ftextr}) to
have a condensed solution. Thus we have the usual scenario of a
continuous phase transition: there is a phase boundary
(\ref{phaseboundinstab}), at which the normal state becomes unstable
and a stable, condensed solution appears.

\subsection{Density equation}

\begin{figure*}[t!]
\centerline{\psfig{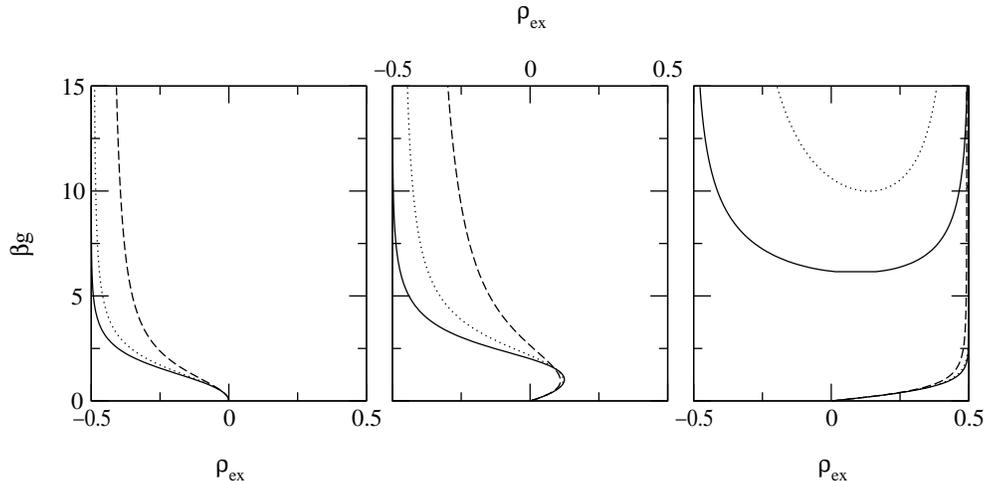}}
\caption{Phase boundaries for $\Delta=0$(left panel), $\Delta=1$(center panel) 
and $\Delta=3$(right panel), and variances $\sigma=0$(solid lines),
$\sigma=0.5$(dotted lines) and $\sigma=1$(dashed lines). For
$\Delta=3$, $\sigma=1$ the upper branch of the phase boundary lies off
the scale, while for $\Delta=3$, $\sigma=0.5$ the lower branch is
indistinguishable from the homogeneous case.  }
\label{phasediagfigure}
\end{figure*}

As well as the mean-field equation (\ref{ftextr}), we need the
equation relating the density $\rho_{ex}$ to the corresponding
chemical potential $\mu_{ex}$. This is obtained from the partition
function in the standard way,
\begin{equation} \label{gendens}
\rho_{ex}=\frac{1}{\beta N}\frac{\partial}{\partial\mu_{ex}}\ln Q.
\end{equation} The asymptotic form for the partition function is
$Q\sim e^{-NS_{0}}$, where $S_{0}$ is the minimal action. Inserting
this asymptotic form in (\ref{gendens}) gives, for the solution
$\psi_{0}(\tau)=\lambda$,
\begin{equation}
\rho_{ex}=|\lambda|^2-\frac{1}{2N}\sum_{n}\frac{\tilde\varepsilon_{n}}{E_{n}}\tanh\left(\beta
E_{n}\right),
\label{ftexdens}
\end{equation} which is the generalization of (\ref{ztexdens}) to
finite temperatures. 

The first term in (\ref{ftexdens}) is the contribution to the
excitation density from the macroscopic electromagnetic field, while
the second term is the contribution from the thermal population of
renormalized electronic excitations. In the absence of a macroscopic
electromagnetic field, $\lambda=0$, both the photon contribution and
the renormalization of the electronic excitations disappear. The
expression (\ref{ftexdens}) is then the familiar form for the
excitation of a set of two-level oscillators.

\section{Phase diagram}
\label{phasediag}

From (\ref{phaseboundinstab}) and (\ref{ftexdens}) we have the
critical temperature for condensation, as a function of the excitation
density, in the homogeneous model:
\begin{equation}
\label{homphasebound}
\beta_{c}g=\frac{4\tanh^{-1}(2\rho_{\mathrm ex})}
{\Delta\pm\sqrt{\Delta^2-8\rho_{\mathrm ex}}}.
\end{equation}  Note that the transition temperature
depends logarithmically on the density, and its scale is set by the
interaction strength $g$. This is in contrast with a model of
propagating, weakly-interacting bosons, where the transition
temperature varies as a power law of the density and its scale is set
by the mass of the bosons.

At low densities, (\ref{homphasebound}) is the phase boundary
separating a population of electronic excitations with energy $E_{0}$
from a population of conventional polaritons with energy
$E_{\mathrm{LPB}}$. To see this, note that such a transition would
occur when the chemical potential for the electronic excitations
reaches $E_{\mathrm{LPB}}$, corresponding to a density
$\rho_{ex}+0.5\approx e^{-\beta_{c}(E_{0}-E_{\mathrm LPB})}$, which is
the low-density limit of (\ref{homphasebound}).

For the inhomogeneous model, we calculate the phase boundary
numerically, assuming the same Gaussian distribution of energies as in
section\ \ref{variationalresults}. We obtain the critical chemical
potential for condensation, $\mu_{c}(\beta_{c})$, by demanding that
(\ref{ftextr}) have a repeated root $\lambda=0$, and then use
(\ref{ftexdens}) to obtain the critical density $\rho_{c}(\beta_{c})$.

In Fig.\ \ref{phasediagfigure} we plot the homogeneous phase
boundaries (\ref{homphasebound}), along with numerical results for the
inhomogeneous model with $\sigma=0.5$ and 1. On resonance, $\Delta=0$,
the transition temperature increases monotonically with density. The
system is always condensed for $\rho_{ex}>0$, because to exceed this
density would require a chemical potential above the center of the
energy distribution of the electronic excitations, and hence above the
bosonic cavity mode. While for $\Delta<0$(not illustrated) the phase
boundary is qualitatively unchanged from the resonant case, for
$\Delta>0$ we find reentrant behavior. This behavior is the result of
the saturable nature of the electronic states. It can be understood by
considering the limits $\rho_{ex}\to\pm 0.5$ when $\Delta \gg 0$. Near
the $\rho_{ex}=-0.5$ limit, the normal state consists of a small
number of electronic excitations, weakly interacting with each other
through the cavity mode. They condense when their density exceeds a
critical value set by the strength of their interaction, which is
determined by $\Delta$ and $g$. Near the $\rho_{ex}=0.5$ limit, the
electronic system is constrained to be fully occupied, and the normal
state consists of a small number of holes in an otherwise completely
excited electronic system. These holes again interact through the
cavity mode, and so the transition occurs when the density of holes,
$0.5-\rho_{ex}$, exceeds a critical value. For $\Delta\to\infty$ the
critical densities of holes and excitons are identical, so the phase
diagram is symmetric about $\rho=0$. For finite $\Delta$, the
interaction is stronger for the holes than for excitons, since they
are nearer in energy to the cavity mode, and so the phase boundary
becomes skewed to the forms shown.

\begin{figure}[t]
\centerline{\psfig{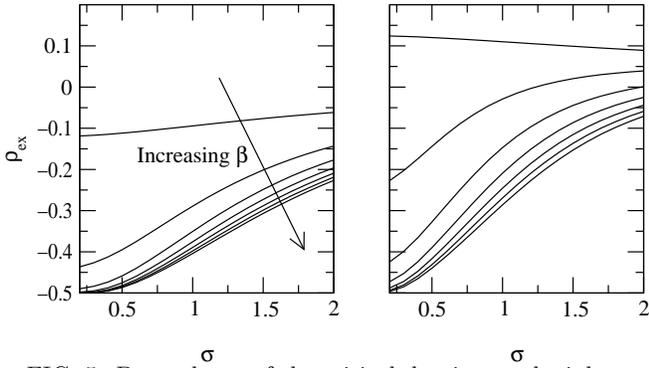}}
\caption{Dependence of the critical density on the inhomogeneous
broadening $\sigma$, for $\Delta=0$(left panel) and $\Delta=1$(right
panel), and $\beta=1$(top curve), 3, 5, 7, 9, 11, 13(lowest curve).  }
\label{phaseboundinhomfigure}
\end{figure}

At temperatures which are high compared with the inhomogeneous
broadening $\sigma g$, thermal fluctuations dominate over the
inhomogeneous broadening. Thus at these temperatures the inhomogeneous
broadening has little effect, as can be seen in Fig.\
\ref{phasediagfigure}. However, at low temperatures the inhomogeneous
broadening suppresses condensation by increasing the energy separation
between the electronic excitations and the photons, collapsing the
phase boundaries towards $\rho_{ex}=0$.  The effects of inhomogeneous
broadening are further illustrated in Fig.\
\ref{phaseboundinhomfigure}, which shows the dependence of the critical density on
$\sigma$ at various temperatures for detunings $\Delta=0$ and
$\Delta=1$.

\section{Excitation energies}
\label{exspec}

In this section, we use (\ref{qpseffinhom}) and (\ref{qpseff0}) to
study the excitation spectrum of the quasi-equilibrium states of the
model (\ref{ham}). The excitation spectra we calculate explain the
form of the phase diagrams in Fig.\ \ref{phasediagfigure}. The excitation spectra of the two
quasi-equilibrium states are different from each other, and also from
the excitation spectrum of a conventional laser. Since the excitation
spectrum is directly related to the optical absorption spectrum of the
cavity, which is an experimentally accessible quantity, these spectra
offer a clear experimental signature of polariton condensation.

The matrix ${\mathcal{G}}^{-1}$, given by (\ref{qpseffinhom}), is the
inverse of the thermal Green's function for the photons. We use the
standard relations\ \cite{hydroflucbook,negorl} between thermal
Green's functions, retarded Green's functions, and excitation spectra,
to extract the latter from (\ref{qpseffinhom}).

\subsection{Homogeneous model}

We begin with the normal state of the homogeneous model. The inverse
of the normal state Green's function contained in (\ref{qpseff0}) can
be written as a sum of simple poles
\begin{equation}
{\mathcal{G}}(\omega_{n})=\frac{C_{+}}{i\omega_{n}+E_{+}}
+\frac{C_{-}}{i\omega_{n}+E_{-}}.
\label{thgreen}
\end{equation} The structure of this Green's function is clear: we
have two excitations, with quasiparticle energies
\begin{displaymath} E_{\pm}+\mu_{\mathrm{ex}}=[(\omega_{c}+E_{g})\pm
g\sqrt{\Delta^2-8\rho_{\mathrm ex}}]/2,\end{displaymath} and
corresponding weights
\begin{displaymath}
C_{\pm}=\pm(2\tilde{\varepsilon}-E_{\pm})/(E_{-}-E_{+}).\end{displaymath}
These normal-state excitations are polaritons in the general sense of
Hopfield\ \cite{hopfpol}: coupled modes involving the linear response
of the electronic system around its equilibrium state. The gap in the
spectrum is increased over the bare detuning $\Delta$ owing to the
dipole coupling between the excitons and the cavity mode.  The
presence of excitation in the ground state, either driven by finite
temperatures or by finite $\mu_{\mathrm ex}$, causes the two polariton
branches to attract. This attraction is due to the decrease in the
polarizability of the electronic states as their population increases
and saturation occurs. It can also be understood in terms of an
angular momentum representation\ \cite{dickemodel} for the collective
states of the electronic system. In such a representation, the
excitation of the electronic states corresponds to the z component of
an angular momentum, while their polarization corresponds to the
raising operator $S_{+}$. Thus the polarizability of the electronic
states is maximized at $\langle S_{z}\rangle=-N/2$.

Since condensation is a phase transition, we expect a qualitatively
different excitation spectrum in the condensed state. From
(\ref{qpseffinhom}) and (\ref{ftextr}), we find for the leading
component of the matrix thermal Green's function
\begin{equation}\label{thermalgf}
{\mathcal{G}}_{11}(i\omega_{n})=\frac{\tilde{\omega}_{c}(\omega^{2}+2g^{2}|\lambda|^{2})-i\omega(\omega^{2}+4E^{2}+2\tilde{\omega}_{c}\tilde{\varepsilon})
}{
(i\omega_{n})^2(i\omega_{n}+\xi)(i\omega_{n}-\xi)(1+\delta_{\omega_{n}}\alpha)},
\end{equation} with
$\xi=\sqrt{(\tilde{\omega}_{c}+2\tilde{\varepsilon})^2+4g^2|\lambda|^2}$. The
interpretation of this Green's function is complicated because, as we
have already mentioned, it describes both quasiparticle excitations
and fluctuations of the order parameter. To rigorously obtain the
quasiparticle spectrum, we should extract the contribution to
(\ref{thermalgf}) from the order parameter fluctuations, and
analytically continue the remainder to obtain the retarded Green's
function. Rather than follow such a procedure, we propose the
following physically appealing if mathematically na\"{\i}ve
interpretation of (\ref{thermalgf}): the Kronecker delta and the
$(i\omega_{n})^2$ terms in the denominator describe the condensate
response, leaving quasiparticle excitations at energies $\pm \xi$. The
$(i\omega_{n})^2$ is clearly associated with the phase mode of the
condensate, discussed in Appendix\ \ref{goldstappendix}, while the
Kronecker delta is related to number fluctuations of the condensate\
\cite{condensate-amplitude-mode}. The excitations at energy $\pm \xi$
are coupled exciton-photon modes in the presence of the macroscopic
electromagnetic field of the condensate. $\xi$ is analogous to the
pair breaking energy in a superconductor: it is the energy required to
extract an exciton-photon complex from the condensate. Note that if we
remove the photon contribution to this energy, by setting
$\tilde{\omega}_{c}=0$, then $\xi$ becomes the familiar expression\
\cite{gelesin} for the energy of an electron-hole pair in the presence
of a classical electromagnetic field at frequency $\mu_{ex}$.

\begin{figure}[t]
\centerline{\psfig{file=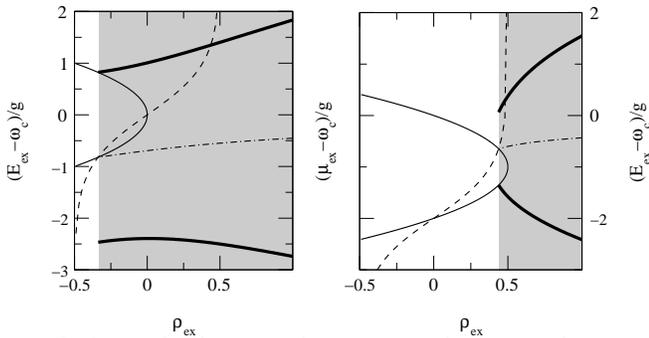,width=8.6cm,clip=true}}
\caption{Excitation energies and chemical potentials as a function of
density for the homogeneous model at $\Delta=0$ (left panel) and
$\Delta=2$ (right panel), both with $g \beta=2$. Thin solid lines:
normal state excitation energies. Thick solid lines: condensed state
excitation energies. Dashed lines: normal state chemical
potential. Dot-dashed lines: condensed state chemical potential. The
shading marks the condensed region for this $\beta$.}
\label{homoexcitationsfigure}
\end{figure}

In Fig.\ \ref{homoexcitationsfigure}, we illustrate the evolution of
the excitation energies of the microcavity with increasing density. To
explain the relationship between the excitation energies and the phase
diagram, we also plot the chemical potentials for the normal and
condensed states on this figure. The left panel of this figure should
be compared to the $g \beta=2$ line of the corresponding phase
diagram, which is the left panel of Fig.\ \ref{phasediagfigure}. When
$\Delta=0$ and $\rho_{\mathrm ex}=-0.5$ the system is in the normal
state. Increasing $\rho_{\mathrm ex}$ populates the electronic
excitations, increasing the chemical potential and decreasing the
polariton splitting. Eventually the chemical potential crosses the
lower polariton branch from below and the system condenses. At the
critical density, the lower polariton branch joins to the phase mode
at the chemical potential, the upper branch joins to the ``pair
breaking'' excitation, and an excitation appears below the chemical
potential. This latter excitation has zero weight at the
transition. It corresponds to an excited state to ground state
transition, where an exciton-photon complex is absorbed into the
condensate. There is no corresponding excitation in the normal state
Green's function, because the ground state of the $N+1$ particle
system($N+1$ excitons) cannot be reached from the excited states of
the $N$ particle system($N-1$ excitons and 1 polariton) by adding a
photon.

The relationship between the excitation spectrum and the phase diagram
is slightly different when the transition occurs for
$\rho_{ex}>0$. For example, in the right panel of Fig.\
\ref{homoexcitationsfigure} the chemical potential crosses the lower
polariton branch at $\rho_{\mathrm ex}=0$ without the condensate
appearing. It is not until the chemical potential crosses the upper
polariton branch that the transition occurs. This can be understood by
considering the signs of the quasiparticle weights $C_{\pm}$. A
positive quasiparticle weight corresponds to absorption of an external
field(a particle-like transition), whereas a negative quasiparticle
weight corresponds to gain(a hole-like transition). For $\rho_{\mathrm
ex}>0$, the lower polariton branch has a negative weight: it has
become hole-like, and must be below the chemical potential for
stability. At the transition it is now this lower branch which joins
to the ``pair forming'' excitation of the condensate, while the upper
branch joins to the phase mode and the ``pair breaking'' excitation
appears above the phase mode.

\subsection{Inhomogeneous model}

Since the inhomogeneous model has a distribution of excitations, we
must study the spectral function $A(\omega)$. $A(\omega)$ is
proportional to the imaginary part of the retarded Green's function,
\begin{equation}\label{specdensity}
A(\omega)=2\Im G^{R}(-\omega+\mu_{ex}).\end{equation} It is
proportional to the optical absorption coefficient of the cavity at
frequency $\omega$, i.e. the imaginary part of the dielectric
susceptibility.

To obtain $A(\omega)$ we require the retarded Green's function
$G^{R}$. In the normal state this is given by the straightforward
analytical continuation \begin{equation}\label{analyticcont}
G^{R}(\omega)=\lim_{\eta\to 0^+}
{\mathcal{G}}(i\omega_{n}=\omega-i\eta).
\end{equation} However, in the condensed state we face the
problem, already mentioned for the homogeneous case, of separating the
order parameter response from the quasiparticle response. We
circumvent this problem by simply assuming that the continuation
(\ref{analyticcont}) of the normal state Green's function also applies
in the condensed state.

Inverting the ${\mathcal{G}}^{-1}$ contained in (\ref{qpseffinhom})
and using (\ref{specdensity}) and (\ref{analyticcont}) expresses
$A(\omega)$ in terms of integrals over the distribution of energies of
the two-level oscillators. We evaluate these integrals in the limit
$\eta \to 0$ by setting $\eta=0$ in the integrands and deforming the
contour of integration around the poles of the integrand on the real
axis. The contribution to the integrals from the detour around the
poles can be performed analytically, leaving a principal value
integral which we evaluate numerically.

\begin{figure*}[t]
\centerline{\psfig{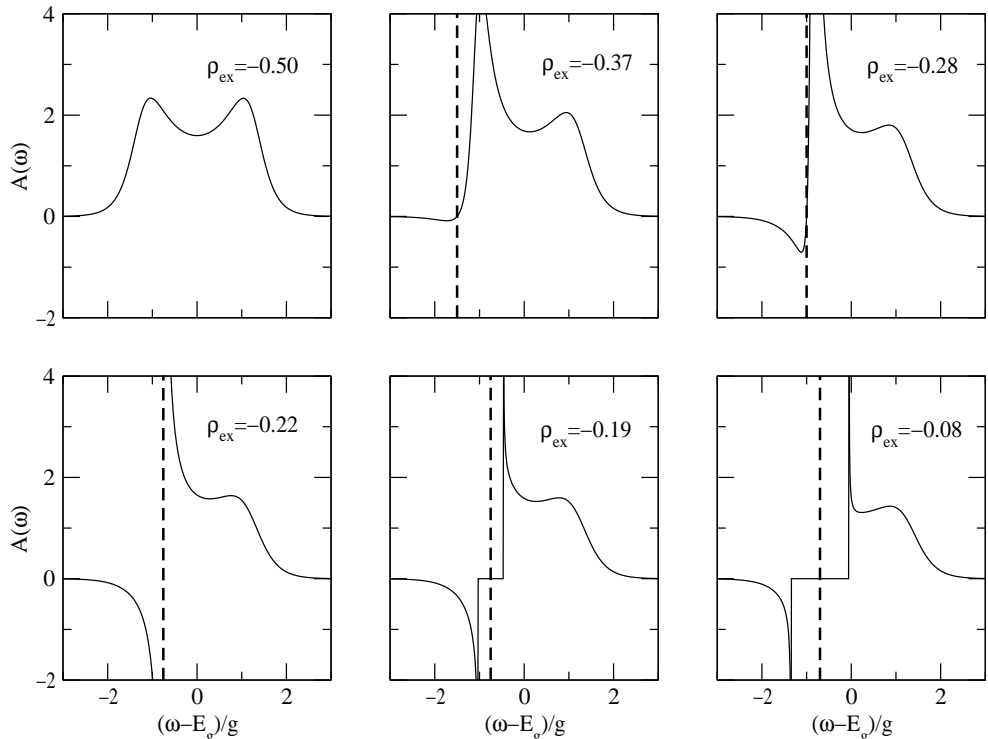}}
\caption{Spectral functions(optical absorption spectra) $A(\omega)$, for $\Delta=0$, $g\beta=2$,
$\sigma=1$ and chemical potentials $(\mu_{ex}-E_{g})/g=-5, -1.5, -1.0,
-0.76, -0.75, -0.70$, increasing from top left to bottom right through
the transition at $(\mu_{ex}-E_{g})/g=-0.76$. The top row of plots are
in the normal state, the bottom left hand plot at the transition and
the remaining plots in the condensed state. The vertical dashed lines
mark the chemical potential.}
\label{inhomexcitationsfigure}
\end{figure*}

Figure\ \ref{inhomexcitationsfigure} shows the evolution of our
calculated absorption spectra, $A(\omega)$, as we increase the density
through the transition, for $g \beta=2$, $\sigma=1$, and
$\Delta=0$. The corresponding chemical potential is marked as the
dashed line. For the empty cavity, $\rho_{ex}=-0.5$, we recover the
absorption spectrum calculated by Houdr\'{e} et al.\
\cite{inhompols}. Comparison with Fig.\ \ref{homoexcitationsfigure}
shows that, for these parameters, the positions of the polariton peaks
are largely unaffected by the inhomogeneous broadening. However, since
the polaritons are now resonant with a significant density of
electronic states they become broadened. Increasing the chemical
potential, but remaining in the normal state, we see the thermal
occupation factors producing gain below the chemical potential and
increased absorption just above. The collapse of the polariton
splitting evident in Fig.\ \ref{homoexcitationsfigure} is hardly
noticeable at these low densities. As the density is increased still
further a pole appears in $A(\omega)$ at the chemical potential; this
marks the onset of condensation. Above the critical density the
coherent cavity field, oscillating at frequency $\mu_{ex}$, produces a
gap of magnitude $4g|\lambda|$ in the spectrum. The peak on the high
energy side of the gap connects smoothly to the upper polariton peak
of the normal state, just as in the homogeneous case. In the
homogeneous case we noted the appearance of an excitation below the
chemical potential as we crossed the transition. This is still present
in the inhomogeneous case, but for the parameters used in
Fig. \ref{inhomexcitationsfigure} it is far too weak to be visible.

\section{Conclusions}
\label{conclusions}

Real microcavities are far more complex than the idealized model
(\ref{ham}). However, like our model, they consist of photons coupled
to electronic excitations which are bosons at low densities, but
reveal their fermionic internal structure at high densities. We have
shown how the polariton condensate may be generalized to allow for the
saturation nonlinearity produced by such fermionic structure. The
saturation nonlinearity can produce (1)a collapse of the splitting
between the peaks in the absorption spectrum of the normal state with
increasing density, (2)a shift of the chemical potential of the
condensate away from the conventional polariton energy, and (3)an
unusual reentrant phase boundary for condensation. 

Experimental work on cavity polaritons has concentrated on
microcavities containing high-quality GaAs quantum wells.  In these
systems, the excitons are weakly-bound, and rather delocalized. Thus,
while the saturation nonlinearity discussed here is present for these
excitations, it will be accompanied by other nonlinearities produced
by the overlap of the wavefunctions of different excitons and the
ionization of excitons\
\cite{cavpol-bleaching,cavpol-bleaching-2}. These effects may well
prevent condensation, but are separated from the saturation
nonlinearity considered here in systems with localized, tightly-bound
excitons. Note also that tightly-bound excitons have a large dipole
coupling $g$, and hence the transition temperature will be larger.

For real examples of localized oscillators, there will be some energy
$E_{m}$ above which delocalized states exist. The picture of a
condensate formed from localized oscillators then only holds when
$E_{m}-\mu_{ex}$ is large compared with $\beta^{-1}$ and $g$. By
considering Fig.\ \ref{homoexcitationsfigure}, we deduce that to
completely realize a reentrant phase diagram like that shown in Fig.\
\ref{phasediagfigure} requires an energy gap $\Delta E$ separating the
localized and delocalized excitations; this gap must be large compared
with $g$ and $\beta^{-1}$. Such a gap could occur in organic
semiconductors\ \cite{organicpol1,organicpol2}. In these systems,
excitons are strongly bound and therefore small(Frenkel). They readily
self-trap on local lattice distortions and on impurities in these,
often highly disordered, materials. An energy gap $\Delta E$ could
exist in inorganic quantum wells if the excitons move in a potential
containing deep, well-separated traps, perhaps associated with
interface islands in narrow quantum wells\
\cite{gammon1,hess,saturosc}.

The disordered quantum wells studied by Hegarty et al.\
\cite{exmobilityedge} provide an example of a system without a gap
separating the localized and delocalized excitations. These systems
show a single inhomogeneously broadened exciton line, unlike the
quantum wells of Refs.\ \onlinecite{gammon1,hess,saturosc}. The
``mobility edge'' $E_{m}$ lies near to the center of the exciton
line. One may be able to form a condensate which does not involve
delocalized excitations using this type of quantum well if the
inhomogeneous broadening is large compared with $g$ and $\beta$ and
the cavity mode is placed low down in the exciton line. The transition
would then occur when the chemical potential is well separated from
$E_{m}$.

The polariton condensate described here is formed from a quasi-thermal
population of electronic excitations which are renormalized by the
coherent photons in the cavity. This renormalization, embodied in the
Bogolubov transformation (\ref{bogt}), produces a gap of magnitude
$4g|\lambda|$ in the absorption spectrum of the condensate. Such
renormalizations, and hence the gap, are absent in conventional
semiconductor lasers\ \cite{sclaserbook}, for which a quasi-thermal
population of the {\emph{bare}} electronic excitations is
assumed. Thus the presence or absence of a gap allows the polariton
condensate to be distinguished from a conventional laser.

In a conventional laser, the renormalization of the electronic
excitations by the photons, and hence the gap, is absent because the
electronic polarization is very heavily damped. The destruction of a
gap by damping is well-known in superconductors, where it is
associated with magnetic impurities. Such impurities suppress the gap,
eventually to zero. The destruction of the gap does not coincide with
the destruction of the order parameter however: near to $T_{c}$ there
is a regime of gapless superconductivity, in which there is an order
parameter but no gap in the single-particle spectrum. This regime
should correspond to the conventional semiconductor laser, although
the actual damping mechanisms will differ.

The non-equilibrium analog\ \cite{noneqstarktheory} of the crossover
illustrated in Fig.\ \ref{homoexcitationsfigure} and Fig.\
\ref{inhomexcitationsfigure}, from a two-peaked polariton spectrum to
a ``Stark triplet'', has been observed experimentally\
\cite{polstarkcross}. In that experiment, the gapped absorption
spectrum is observed simultaneously with the excitation pulse. Thus
there is no thermalization involved in producing the gapped
spectrum. It is the result of coherence in the excitation pulse,
rather than the spontaneous coherence of condensation. Nonetheless,
these experiments demonstrate the renormalization of the electronic
states that is essential in the polariton condensate.

To reach the quasi-equilibrium regime we have described requires a
system where the polariton lifetime is long compared with the time
required to reach thermal equilibrium at a fixed number of
polaritons. Current semiconductor microcavities have lifetimes for the
photons, and hence the polaritons, of the order of
picoseconds. Finding an exciton system which thermalizes on this
timescale may be difficult. However, there seems no reason to suppose
it is impossible, particularly beyond the linear regime, since
nonlinearities can enhance relaxation\ \cite{senbloch}. Furthermore,
microcavities are available with lifetimes far greater than
picoseconds.  For example, silica microspheres have confined modes
with lifetimes of microseconds\ \cite{silicaresonatorq}.

\section{Acknowledgments}

This work was supported by the Engineering and Physical Sciences
Research Council, U.K.

\appendix
\section{The phase mode}
\label{goldstappendix}

In this appendix we investigate the zero eigenvalue of
${\mathcal{G}}^{-1}$ that appeared while studying the stability of the
condensate in the homogeneous case. We first prove that the zero is
also present in the inhomogeneous model, and that it describes phase
fluctuations of the condensate. It is thus the Goldstone mode
reflecting the degeneracy of the ground state with respect to the
phase of the order parameter. We then argue that the zero eigenvalue
does not contribute to the free energy density in the thermodynamic
limit. Although the physics we discuss in this appendix is well
understood in general, it is particularly transparent in our simple
model.

We note that, at $\omega=0$, $K_{1}$ is real and positive. The
eigenvalues of ${\mathcal{G}}^{-1}$ are then $K_{1}\pm|K_{2}|$. Since
from the explicit forms of $K_{1}$, $K_{2}$ and the extremal equation
(\ref{ftextr}) we have $|K_{2}|=K_{1}$, as required in general by the
Hugenholtz-Pines relation\ \cite{pybook,bosehugpines},
${\mathcal{G}}^{-1}$ has a zero eigenvalue.

To illustrate that the zero eigenvalue is the phase mode of the
condensate, note that since $\arg K_{2}=2\arg\lambda=2\phi$ we can
write
\begin{displaymath} {\mathcal{G}}^{-1}\propto \left(
\begin{array}{cc} 1 & e^{2 i \phi} \\ e^{-2 i \phi} & 1
\end{array}\right). \end{displaymath} The eigenvector of this matrix
with zero eigenvalue is perpendicular in the complex plane to the
order parameter $\lambda$.

Since we are considering a broken symmetry system, we should not
include states with different phases of the order parameter when
calculating the partition function. Thus on physical grounds, we
should discard the zero mode when computing the partition function.

A formal approach which allows calculations in the presence of the
zero eigenvalue is to introduce symmetry breaking terms which are
taken to zero after the thermodynamic limit. This is the standard
method for applying statistical mechanics to broken symmetry systems\
\cite{negorl}. The appropriate symmetry breaking terms for a Bose
condensed system pin the phase of the order parameter. They are
sources and sinks for the photons, and appear in the effective action
$S_{\mathrm eff}$ as
$\frac{1}{\sqrt{N}}\left(\bar{\psi}J+\bar{J}\psi\right).$ These terms
do not contribute directly to (\ref{qpseffinhom}), but appear as a
source term in (\ref{ftextr}). The original zero eigenvalue of
${\mathcal{G}}^{-1}$ is now $K_{1}-|K_{2}|=-J/(\psi_0\sqrt{N})$. Since
for the equilibrium solution we must have $\phi-\arg J = \pi$, the
contribution of the original zero eigenvalue to the free energy
density is proportional to
\begin{displaymath} \lim_{J\to 0}\lim_{N\to\infty}\frac{1}{N}\ln
\left(\frac{|J|}{|\lambda|\sqrt{N}}\right) = 0.
\end{displaymath}

\end{document}